\pdfoutput=1
\documentclass[12pt,preprint,epsf]{aastex} 

\usepackage{epstopdf}

\shorttitle{Solar Panel Evaluation}
\shortauthors{Krisciunas} 

\begin{document}
\received{8 June 2022}

\title{Including Atmospheric Extinction in a Performance
Evaluation of a Fixed Grid of Solar Panels}   
\author{
Kevin Krisciunas\altaffilmark{1,2}
}
\altaffiltext{1}{Texas A. \& M. University, Department of Physics \& Astronomy,
  4242 TAMU, College Station, TX 77843; {krisciunas@physics.tamu.edu} }

\altaffiltext{2}{George P. and Cynthia Woods Mitchell Institute for Fundamental 
Physics \& Astronomy }

\begin{abstract} 
We characterize the performance of a fixed grid of solar panels on the basis
of data taken under clear sky conditions over 12 months.  
We confirm that the power output is linearly proportional to 
cos($\theta$), where $\theta$ is the angular difference of direction toward 
the Sun and the vector perpendicular to the panels.  In order to confirm 
this we applied methods from astronomical photometry reduction. 
From late March through August we find that the median 
effective atmospheric extinction term is 0.145 mag/airmass.  From October
to mid-March the median extinction term is 0.081 mag/airmass. The 
proportionality ``constant'' scaling cos($\theta$) appears to be seasonally 
dependent, with the smallest scaling factors occurring when the extinction 
term is largest.  Finally, we find that extinction-corrected power 
often underperforms the linear relationship late in the morning or early in 
the afternoon.  This is most likely because the efficiency of solar panels 
depends on their operating temperature, and the panel temperature increases 
over the course of time on a sunny day.
\end{abstract}


\section{Introduction}

On 26 April 2021 Texas Green Energy installed two sets of solar 
panels on my house in College Station, Texas (latitude  N 30\arcdeg ~33\arcmin
~43\arcsec, longitude W 96\arcdeg ~16\arcmin ~8\arcsec, elevation 85 m).\footnote[3]
{Engineered in Germany by Q.CELLS, the panel model is 
Q.PEAK DUO BLK-G6+ 330-345.  Basic information on photovoltaic cells
can be found here: 
https://www.energy.gov/eere/solar/articles/pv-cells-101-primer-solar-photovoltaic-cell.
Also useful is this website: 
https://sciencing.com/effect-wavelength-photovoltaic-cells-6957.html} 
Six panels are mounted on the upper roof, and seven panels on the lower roof 
(Fig. \ref{fig:house_panels}). After engineers, technicians, and city 
inspectors were done with their work, the system went live on 8 June.
Over the course of one year the system has generated 6516 
kilowatt-hours (kWh) of power, or an average of 18 kWh per day.

Solar panels produce electric power by means of the photovoltaic effect. 
The photoelectric effect, on the other hand, involves the ejection of 
electrons from a conducting plate when light shines on it.  However, the 
photovoltaic effect takes place at the boundary of two semiconducting 
plates.  Electrons are not ejected.  They accumulate along the boundary 
of the semiconducting plates and create a voltage.  When the plates are 
connected with a wire, a current flows in the wire.

Our solar panels contain crystalline silicon as the semiconducting 
material. One key parameter is the ``band gap energy'' of crystalline 
silicon.  Only photons with energy greater 1.11 electron volts can dislodge 
electrons from silicon atoms and send them into the conduction band between 
the two semiconducting plates.  The longest wavelength light usable by the 
solar panels is near-infrared light at a wavelength of 1.1 microns.  Much 
shorter wavelength photons, with energy greater than 3 eV, send electrons 
out of the conduction band, rendering them unable to do work.  The 
corresponding wavelength is about 413 nanometers (corresponding to violet 
light). Thus, solar panels that use silicon as the semiconducting material 
convert sunlight with wavelengths ranging from 0.413 to 1.1 microns into 
electric power.  The author does not know of a ``response curve'' of solar 
panels as a function of wavelength, but the mean wavelength of sensitivity 
is about 760 nm.  This is between the effective wavelengths of the 
photometric $R$- and $I$-bands at 650 and 800 nm, respectively (or the 
corresponding Sloan $r$- and $i$-bands).

According to the installers, our panels face azimuth 135\arcdeg (i.e., 
southeast). The pitch of the upper set of panels was measured by them
to be 24\arcdeg.  Later we did our own measurements using a phone application
called Measure and found that the
upper set of panels is pitched 23.3\arcdeg ~to the horizontal, while the lower
set of panels is pitched 20.2\arcdeg ~to the horizontal.  We adopt the average
value (21.75\arcdeg) for subsequent analysis.  

The vector perpendicular to the panels intersects the celestial sphere at 
azimuth 135\arcdeg \hspace{0.3 mm}and elevation angle 68.25\arcdeg 
\hspace{0.3 mm} in the horizon system of celestial 
coordinates.\footnote[4]{Basic information on the celestial sphere can be 
found online at http://people.tamu.edu/~kevinkrisciunas/cel\_sphere.html .} 
Let us call this position P. Its elevation angle and azimuth has an 
uncertainty of perhaps $\pm$1 deg in each coordinate. Such uncertainty will 
not have a significant effect on the main results presented here.

Now, just as the sky is brighter and brighter during morning twilight, and 
the world outside is not pitch dark on a cloudy day, 
solar panels can generate some power even if no sunlight is directly 
incident upon them. However, since light from the Sun's photosphere is so much 
brighter than sunlight scattered by the various components of the Earth's 
atmosphere, here we try to characterize the power output from the solar 
panels solely from direct sunlight.  To characterize the performance of
the solar panels we will need to calculate the elevation angle and azimuth
of the Sun at multiple times on any given day.

For an object like the Sun that transits the celestial meridian
south of the zenith for an observer in the northern hemisphere
situated north of the Tropic of Cancer,\footnote[5]{In Hilo, Hawaii, 
for example, at latitude +19.7\arcdeg, the Sun transits the celestial
meridian {\em north} of the zenith for about two months each year starting
on May 18th.} the maximum elevation angle above the horizon will be:

\begin{equation}
h_{max} \; = \; [90\arcdeg \; - \; \phi] \; + \delta _{\odot} \; ,
\label{eq:hmax}
\end{equation}

\parindent = 0 mm

where $\phi$ is the latitude of the observer and $\delta _ {\odot}$ 
is the declination of the Sun.  Over the course of the year $h_{max}$
has a range of 47\arcdeg (twice the tilt of the Earth's axis of
rotation to the plane of its orbit).


\parindent = 9 mm

Let $\theta$ be the angular distance between position P mentioned above and 
the position of the Sun on the celestial sphere.  Of course, due to the 
rotation of the Earth, the elevation angle and azimuth of the Sun change 
continuously, so $\theta$ changes continuously.

Fig. \ref{fig:raw_data} is a plot of raw data obtained from 9 June 
to 21 August 2021 and tabulated in Table \ref{tab:data}.
The reader will notice that the data are not symmetric. Starting about 
15:15 CDT a summer day, the lower panels are more and more in shadow.
We will not consider any further those measurements in Table 
\ref{tab:data} taken after 15:15 local time.  Also, one should note that
the data obtained from 7 to 10 AM Central Daylight Time show very little
scatter from day to day.  By noontime something else is happening.  We
believe it is related to the efficiency of the solar panels as a function
of operating temperature (see Discussion below).

Table \ref{tab:moredata} contains raw data from three particular days 
that will be discussed below.  Raw data for 17 other days analyzed
in this paper can be obtained from the author by request. 

In Fig. \ref{fig:3d_raw_data} we plot one year's worth of raw data in a 3D 
projection plot.  A few data points were excluded: those taken when there was 
frost on the solar panels, and those taken when some of the panels were in shadow.  
Altogether, 333 data points are plotted.

According to the specifications on solar panels, their output varies 
proportionally to cos($\theta$).  This can be demonstrated by having
a set of solar panels that can track the Sun, compensating
for the rotation of the Earth.  The user would also
want to be able to aim the panels any number of degrees to the right or 
left, and above or below the direction to the Sun.  This is beyond
the capabilities of our system.  That is just as well, as the local
Home Owners Association would veto a plan to install such a system.

We will be able to demonstrate the cosine law. But first we must 
discuss some basics of astronomical photometry.

\section{Astronomical Photometry}

In this section we present an example that shows, in effect, if you observe
an astronomical source and plot the logarithm of the number of photons 
per second detected vs. a measure of the path length through the Earth's
atmosphere, you can fit the data with a straight line.  The slope of that
line is called the atmospheric extinction coefficient.

Astronomical photometry dates back to the time of Hipparchus (ca. 150 B.C.), 
who classified the brightness of the stars using magnitude bins.  The 
brightest stars in the sky were defined to be magnitude 1 (i.e., stars of the 
first class), while the faintest stars visible to the unaided eye are
magnitude five or six.  Using a nineteenth century formulation of 
magnitudes, a first magnitude star gives us exactly 100 times as many 
photons per second as a sixth magnitude star.  In the nineteenth century 
Karl August Steinheil and Gustav Theodor Fechner demonstrated that the 
impression we have of the brightness of a light source is proportional to 
the logarithm of the light intensity.  This relationship holds for hearing 
and for taste and is known as the Weber-Fechner psychophysical law 
\citep[][pp. 70-76]{her84}.

If we measure $f_1$ photons per second from a bright star and $f_2$ 
from a second (fainter) star, then by definition the {\em difference}
of their apparent magnitudes is $-$2.5 log$_{10}$ ($f_1/f_2$) = $\Delta$m.  
Since $f_1 / f_2 > 1.0$, $\Delta$m is negative.  Similarly, 
$a^{\Delta m}$ = $f_1/f_2$, where $a$ is 
equal to the fifth root of 100, or approximately 2.51186.  

Consider a schematic diagram of a plane parallel atmosphere (Fig. 
\ref{fig:slab}).  A star at zenith angle $z$ is observed through a longer 
path length of atmosphere than it would be at the zenith.  (The zenith angle
is 90\arcdeg ~minus the elevation angle.) For $z < 
60\arcdeg$ that path length, divided by the path length towards the zenith, 
is simply sec($z$). We call the ratio of these path lengths the ``air mass'' 
(X), which is a unitless parameter.  It is {\em not} a column density of 
molecules and atoms measured through the atmosphere.  For $z < 60\arcdeg$,
X = sec($z$) is a good approximation.  This is calculated using spherical
trigonometry.

For zenith angles greater than 60\arcdeg (i.e., air mass greater than 2.0) 
the plane parallel approximation of the atmosphere breaks down.  After all, 
the atmosphere is essentially a large number of thin spherical shells
whose density decreases with altitude.  Also, for objects viewed near the 
horizon we have to consider atmospheric refraction.  These considerations are 
beyond the scope of the present paper.  For our purposes here we may use the 
following formula from \citet {Har62}:

\begin{equation}
\rm{X} \; = \; \rm{sec} \; z - 0.0018167 \; (\rm {sec} \; z - 1) 
       - 0.002875 \;  (\rm {sec} \; z - 1)^2
       - 0.0008083 \; (\rm {sec} \; z - 1)^3 \; .
\label{eq:hardie}
\end{equation}

\parindent = 0 mm

We note that this formula breaks down for zenith angles greater than 87\arcdeg.

\parindent = 9 mm

Nowadays we take celestial images using a telescope and solid state 
cryogenically-cooled charge-coupled device (CCD), using a number of 
photometric filters, such as a blue filter called $B$ and a yellow-green 
filter called $V$.  These two filters have transmission curves that peak at 
about 440 and 550 nm, respectively.  Instrumental magnitudes can be 
determined by displaying the images with SAO Image DS9 and doing aperture 
photometry in the {\sc iraf} environment.\footnote[6]{{\sc iraf} is 
distributed by the National Optical Astronomy Observatory, which is 
operated by the Association of Universities for Research in Astronomy, 
Inc., under cooperative agreement with the National Science Foundation 
(NSF).}

How does one reduce the data?  Experts give conflicting advice. As my 
colleague George Wallerstein (1930-2021) used to say, ``Four astronomers, 
five opinions.'' For simplicity's sake, using software like {\sc iraf} one 
designates the radius in pixels of an aperture, centers one star in the 
aperture, and determines the number of digital counts above the sky level.  
{\sc iraf} calculates {\em minus} 2.5 log$_{10}$ of those counts and adds 
2.5 log$_{10}$ of the integration time in seconds.\footnote[7]{{\sc iraf} 
also adds an arbitrary constant here, defaulted to 25.  This is to give 
instrumental magnitudes that are positive and comparable to what the 
eventual reduced values become.} This gives an instrumental magnitude
adjusted to an integration time of one second. From the telescope operating 
system the image file headers contain the air mass values.

Say one has taken $B$-band and $V$-band images of fields of standard stars, 
such as those of \citet{lan92,lan07}. One ends up with instrumental $b$ and 
$v$ magnitudes.  The simplest transformation of the instrumental $v$ magnitudes 
and instrumental $b-v$ colors to standardized $V$-band values in the 
photometric system of Landolt is:

\begin{equation}
V \; = \; v \; -\; k_v X + CT_v (b-v) \; + \; \zeta _v, 
\label{eq:trans}
\end{equation}

\parindent = 0 mm

where: $k_v$ is the $V$-band atmospheric extinction coefficient; CT$_v$ 
is a color term used to correct for differences of transmission as a 
function of wavelength of the filters used by the observer and the ones 
used by Landolt; and $\zeta _v$ is a photometric zero point (or, simply, 
the Y-intercept of some regression line).  If the stars range in color 
from very blue to very red and/or the observations were made over a wide 
range of air mass, one may have to add second order terms to Eq. 
\ref{eq:trans}.

\parindent = 9 mm

Using {\sc iraf} or equivalent {\sc fortran} or {\sc python} code, one can 
solve simultaneously for the extinction term, color term, and zero point. 
(We are effectively fitting the best plane to a three dimensional set of 
data.) If the color term is known robustly, then one can take the difference of 
the color-corrected instrumental $v$ magnitudes and Landolt's standardized 
values and plot these magnitude differences vs. the air mass values to
obtain the extinction term and the photometric zero point.

Fig. \ref{fig:nov26} shows an example of determining the atmospheric $V$-band 
extinction coefficient, using data taken by us with the 0.9-m telescope at 
Cerro Tololo Inter-American Observatory (CTIO) on 26 November 2005 (UT). 
We plot the color-corrected instrumental magnitudes ($v + CT_v (b-v)$) minus
the catalogue magnitudes ($V$) vs. the air mass values.  The slope gives
the $V$-band extinction coefficient $k_v$.

If the sky is clear and 
has stable transparency, one can convert all of the raw measurements taken on 
a given night to standardized, publishable values, with random errors of 
$\pm$ 0.02 magnitudes or better (roughly 2 percent). In Fig. \ref{fig:nov26} 
we find a $V$-band extinction coefficient of 0.164 $\pm$ 0.005 mag/airmass 
and a root-mean-square (RMS) residual of $\pm$ 0.013 mag, after eliminating 
one 5-$\sigma$ outlier from the regression.  This may be compared to the 
median $V$-band extinction of 0.154 mag/airmass from 29 nights of 
observations by us at CTIO and Las Campanas Observatory (LCO) from 21 April 
2001 through 21 December 2012.  In our experience, under clear sky conditions 
the extinction values in a particular photometric band at a particular site 
can vary from night to night at least $\pm$50 percent compared to the mean or 
median value.  

\section{Demonstrating the Cosine Law}

One records the power produced by the solar panel system (from a digital 
display) over a wide range of the Sun's elevation angle under clear sky 
conditions. One then proceeds to calculate the elevation angle and azimuth of 
the Sun for each observation time using the right ascension and declination 
of the Sun.  One can look up $\alpha _{\odot}$ and $\delta _{\odot}$ in the 
annual volume of the {\em Astronomical Almanac} and interpolate to local 
noontime for any given day.  Or one can use the method of \citet{vanF_Pul79} 
to obtain $\alpha _{\odot}$ and $\delta _{\odot}$ to the nearest arc 
minute.\footnote[8]{A table of the Sun's coordinates for each day of this 
year can be found here: 
http://people.tamu.edu/~kevinkrisciunas/ra\_dec\_sun\_2022.html}

Using a program written by us, one can set the date, latitude and 
longitude of the site, and the right ascension and declination of the Sun 
to calculate the Sun's azimuth, elevation angle, and air mass for all the 
corresponding times of day.  Then, one calculates the values of angle 
$\theta$, the angular distance between point P (azimuth 135\arcdeg 
\hspace{0.3 mm}, elevation angle 68.25\arcdeg) and the azimuth and 
elevation angle of the Sun at the times in question.  A plot of the 
measured power ($P_{meas}$) vs. cos ($\theta$) from 16 June 2021 is found 
in the top half of Fig. \ref{fig:jun16}.  It is encouraging that the data 
show a linear relation, but it is clear that a linear fit does not pass 
through the origin. It certainly should do so, because the panels begin 
generating significant power shortly after sunrise once the Sun's light 
is incident on the panels.

We can correct the measured values of the power to the equivalent values
we would have obtained if the Sun were at the zenith.  Consider that the Sun 
observed at zenith angle $z$ is dimmer than it would be at the zenith.  (See Fig.
\ref{fig:slab}.)  To correct the observed power to what we would measure
with the Sun at the zenith we must multiply the observed power by 

\parindent = 0 mm


\begin {center}
$10^{(0.4 \times {\rm extinction \; term} \times {\rm path \; length \; difference})} = $
$2.51186^{k_{\lambda} (X_{\odot} \; - 1)} \;$ .
\end{center}

The extinction term $k_{\lambda}$ is the effective extinction in magnitudes per
airmass integrated over the wavelength range at which the solar panels operate
(0.4 to 1.1 microns).

\parindent = 9 mm

Thus, the power we would measure with the solar panels if the Sun were at the 
zenith (airmass 1) is related to the power when the Sun's airmass 
is $X_{\odot}$ as follows:

\begin{equation}
P_{extcorr} \; = \; 2.51186^{k_{\lambda} (X_{\odot} \; - 1)} \; \times \; P_{meas} \; .
\label{eq:ext_corr}
\end{equation}

\parindent = 0 mm

For the physicist or engineer who is confused or horrified by an encounter
with the stellar magnitude system, a relation equivalent to Eq. \ref{eq:ext_corr} 
can be written as follows:

\begin{equation}
P_{extcorr} \; = \; e^{\alpha (X_{\odot} \; - 1)} \; \times \; P_{meas} \; ,
\label{eq:exp_corr}
\end{equation}

where $\alpha = 0.92103 \; k_ {\lambda}$. 

\parindent = 9 mm

Consider making a plot like the top half of Fig. \ref{fig:jun16}, but 
scale the measured values of the power by the factor given in Eq. 
\ref{eq:ext_corr}, with $k_{\lambda}$ = 0.10 mag/airmass.  Fit a line to 
the data and keep track of the Y-intercept of the plot (and its 
uncertainty).  Do this again for trial extinction values of 0.12, 0.14, 
0.16, and 0.18, for example. Then plot those Y-intercepts vs. the trial 
extinction values and derive the extinction value that would give a 
Y-intercept of 0.0.  In the bottom part of Fig. \ref{fig:jun16} we see 
that an extinction term of 0.130 mag/airmass works well for 16 June 2021.

We note that the first point of the day on 16 June was taken when the Sun's 
elevation angle was 3.03\arcdeg (zenith angle 86.97\arcdeg) and its airmass 
(from Eq. \ref{eq:hardie}) was 13.311.  With an extinction term of 0.130 
mag/airmass, the measured power of 137 Watts implies that the solar panels 
would have produced 598 Watts (4.37 times as much) if the Sun had been at 
the zenith and if the solar panels had been tilted to keep angle $\theta$ the 
same (80.09\arcdeg).

Alternatively, we could try a range of parameters $\alpha$ in Eq. 
\ref{eq:exp_corr} to determine the value that gives a regression line of 
$P_{extcorr}$ vs. cos($\theta$) passing through the origin, as we did  
using $k_{\lambda}$ as the extinction term.

In Table \ref{tab:extvals} we give the derived values of the extinction term 
$k_{\lambda}$ obtained over the past year.  The values are plotted in Fig. 
\ref{fig:extvals}. In spring and summer (26 March to 21 August 2022) our data 
indicate a median extinction term of 0.145 mag/airmass, with a standard 
deviation of the distribution of $\sigma _x = \pm$ 0.028.  In fall and winter 
we find a median extinction term of 0.081 mag/airmass, with a standard 
deviation of the distribution of $\sigma _x = \pm$ 0.020.  Roughly speaking, 
the extinction term at our location in the fall and winter is half as large, 
on average, as in the spring and summer.  Table \ref{tab:extvals} also gives 
the slopes of the $P_{extcorr}$ vs. cos($\theta$) diagrams.

In Fig. \ref{fig:slope} we show data obtained on 22 January and 28 May 2022. 
(The raw data are given in Table \ref{tab:moredata}.) One can clearly see 
that on these two occasions the power from the solar panels is linearly 
proportional to cos($\theta$), but the proportionality ``constant'' is not a 
constant. It depends on the operating temperature of the panels and the 
transmission of the Earth's atmosphere.

In Fig. \ref{fig:corr} we plot the {\em slopes} of the $P_{extcorr}$ vs. cos 
($\theta$) diagrams vs. the extinction terms.  The data points are color 
coded and correspond to three ranges of time: January to April, May to August, 
and October through December.


\section{Discussion}

In Fig. \ref{fig:extinction} we plot the atmospheric extinction values 
obtained by us on 29 nights at CTIO (elevation 2200 m) and LCO (2380 m) from 
21 April 2004 through 21 December 2012.  The filters used were Johnson $U, B, 
V$, Cousins $R$ and $I$, and the somewhat narrower filters Sloan $u, r$, and 
$i$.  The median extinction values were 0.510 mag/airmass for $U$ or $u$, 
0.267 for $B$, 0.154 for $V$, 0.113 for $R$ or $r$, and 0.065 for $I$ or $i$.  
Table 4 of \citet{kri_etal17} gives mean extinctions derived from 
observations of 2004 to 2009 with the 1-m Swope telescope at Las Campanas as 
part of the Carnegie Supernova Project (CSP-I). We found mean extinction 
values of $k_u$ = 0.511, $k_B$ = 0.242, $k_V$ = 0.144, $k_r$ = 
0.103, and $k_i$ = 0.059 mag/airmass. 


For comparison we consider values of the $V$-band extinction 
measured essentially at sea level. At the nearby Texas A\&M Physics 
Observatory during the colder months the $V$-band extinction values 
range from 0.2 to 0.3 mag/airmass (Carona 2021, private 
communication). From nine nights of observations near Los Gatos, 
California (elevation 80 m), from October 1981 to March 1982 we measured 
mean $V$-band extinction of $k_v$ = 0.341 $\pm$ 0.050 mag/airmass. At 75 
m elevation near Keaau, Hawaii, on 19 December 1989 (UT) we measured 
$k_v$ = 0.341 $\pm$ 0.020 mag/airmass on a night that was affected by 
some volcanic haze.  On the next night (clear, but with strong 
scintillation) we measured $k_v$ = 0.187 $\pm$ 0.046 mag/airmass 
\citep{kri90}.

Let us assert that a sensible estimate of the mean $V$-band extinction at sea 
level under clear sky conditions is 0.30 $\pm$ 0.06 mag/airmass. At CTIO and 
LCO the median or mean extinction is 0.15 $\pm$ 0.01 mag/airmass. Thus, the 
$V$-band extinction at sea level is 2.0 times that at CTIO and LCO.  By 
interpolating the data in Fig. \ref{fig:extinction} we estimate that at a 
wavelength of 0.76 microns (the middle of the range of wavelengths of light 
usable by the solar panels) the average atmospheric extinction at the 
elevations of Las Campanas and Cerro Tololo is about 0.075 mag/airmass. At 
sea level the extinction at 0.76 microns would then be about 2.0 $\times$ 
0.075 $\approx$ 0.15 mag/airmass.  This may be compared with the median 
extinction value of 0.145 mag/airmass from data obtained from mid-March 
through August given in Table \ref{tab:extvals} and shown in Fig. 
\ref{fig:extvals}.  Recall that the median extinction term derived from solar 
panel data taken in October through mid-March was 0.081 mag/airmass.

Finally, we consider the efficiency of the solar panels as a function of
operating temperature.
According to the technical specifications of our panels, the normal operating
temperature is 43 $\pm$ 3 deg C (109 deg F).  The temperature coefficient
$\gamma$ is $-$0.36 percent/deg K. This means that for every degree the panels
are hotter than the normal operating temperature, their efficiency decreases
by 0.36 percent.

In Fig. \ref{fig:oct30} we see an example of how this plays out.  The Sun 
rises and gets higher in the sky.  The power corrected to Sun at the 
zenith increases linearly proportional to cos ($\theta$) until the 
minimum angle $\theta$ occurs, then the power begins to decrease.  On 
that particular day between 13:28 and 14:00 CDT the panels began to 
underperform the regression line fit to data obtained earlier, by about 
ten percent.

The technical specifications from the manufacturer of our solar panels 
indicate that the efficiency of the panels will degrade over their 25 year 
lifetime: ``At least 98 percent of nominal power during the first year. Thereafter 
maximum 0.54 percent degradation per year. At least 93.1 percent of nominal 
power up to 10 years. At least 85 percent of nominal power up to 25 years.''
The analytical method and examples given here would allow the owner of 
a system of solar panels to evaluate their performance now and in the future.

\acknowledgments

We thank Don Carona, Nick Suntzeff, and Peter Nugent for useful discussions.
We also thank Steven Boada for Python programming advice.


\clearpage


\begin{deluxetable}{crrrrrc}
\tabletypesize{\scriptsize}
\tablecolumns{7}
\tablewidth{0pc}
\tablecaption{Raw Data - Part I\tablenotemark{a}\label{tab:data}}
\tablehead{ \colhead{Date} &
\colhead{CDT} &
\colhead{P$_{meas}$} &
\colhead{AZ$_{\odot}$} &
\colhead{EL$_{\odot}$} &
\colhead{X$_{\odot}$} &
\colhead{$\theta$} }
\startdata
 9 Jun 2021 & 16:07 & 1779 & 268.19 & 53.13 & 1.2499 & 53.13 \\
        & 17:10 &  863 & 275.76 & 39.55 & 1.5703 & 68.28 \\
\hline
10 Jun 2021 & 12:31 & 3859 & 118.62 & 75.78 & 1.0316 &  9.07 \\
        & 13:38 & 3545 & 201.83 & 81.95 & 1.0099 & 19.96 \\
        & 14:51 & 3266 & 253.24 & 69.82 & 1.0654 & 36.25 \\
        & 16:16 & 1322 & 269.47 & 51.26 & 1.2820 & 55.80 \\
        & 17:10 &  680 & 275.83 & 39.63 & 1.5679 & 68.21 \\
\hline
13 Jun 2021\tablenotemark{b} &  8:19 & 1579 &  75.36 & 22.35 & 2.6153 &  58.24 \\
        &  9:20 & 2459 &  81.73 & 35.25 & 1.7327 &  44.19 \\
        & 10:06 & 2885 &  86.77 & 45.12 & 1.4112 &  33.66 \\
        & 10:55 & 3278 &  93.01 & 55.70 & 1.2106 &  22.71 \\
        & 11:48 & 3744 & 102.65 & 67.03 & 1.0862 &  12.22 \\   
        & 12:24 & 3909 & 114.29 & 74.41 & 1.0382 &   8.96 \\
        & 13:30 & 3905 & 187.75 & 82.61 & 1.0084 &  18.21 \\
\hline
14 Jun 2021 &  7:23 &  513 &  69.22 & 10.82 & 5.2009 &  71.11 \\
        &  7:44 &  900 &  71.55 & 15.09 & 3.7950 &  66.31 \\
        &  8:04 & 1240 &  73.70 & 19.21 & 3.0172 &  61.73 \\
        &  8:24 & 1539 &  75.81 & 23.37 & 2.5086 &  57.14 \\
        &  8:41 & 1812 &  77.59 & 26.94 & 2.1993 &  53.22 \\
        &  9:05 & 2131 &  80.08 & 32.02 & 1.8858 &  47.70 \\
        &  9:48 & 2595 &  84.66 & 41.22 & 1.5176 &  37.82 \\
        & 10:10 & 2801 &  87.14 & 45.96 & 1.3912 &  32.81 \\
        & 11:03 & 3187 &  94.07 & 57.39 & 1.1872 &  21.05 \\
\hline
15 Jun 2021 & 16:46 &  822 & 273.37 & 45.11 & 1.4114 & 62.45  \\ 
        & 17:18 &  586 & 276.91 & 38.23 & 1.6158 & 69.78 \\
        & 17:53 &  363 & 280.59 & 30.77 & 1.9547 & 77.73 \\
\hline
16 Jun 2021 &  6:45 &  137 &  64.53 &  3.03 &  13.311 & 80.09 \\
        &  7:06 &  402 &  67.04 &  7.17 &  7.7826 & 75.34 \\
        &  7:32 &  840 &  70.01 & 12.39 &  4.5764 & 69.42 \\
        &  7:52 & 1181 &  72.20 & 16.47 &  3.4907 & 64.85 \\
        &  8:03 & 1375 &  73.38 & 18.74 &  3.0883 & 62.33 \\
        &  8:31 & 1823 &  76.32 & 24.57 &  2.3943 & 55.90 \\
        &  9:08 & 2225 &  80.16 & 32.39 &  1.8668 & 47.38 \\
        & 10:03 & 2923 &  86.08 & 44.17 &  1.4352 & 34.78 \\
        & 10:30 & 3152 &  89.27 & 49.99 &  1.3056 & 28.68 \\
        & 11:18 & 3412 &  96.10 & 60.33 &  1.1508 & 18.25 \\
        & 11:54 & 3531 & 103.43 & 67.59 &  1.0786 & 11.73 \\
        & 12:28 & 3584 & 114.98 & 74.94 &  1.0356 &  8.99 \\
        & 13:03 & 3827 & 141.54 & 80.98 &  1.0125 & 12.68 \\
\hline
17 Jun 2021 &  8:17 & 1580 &  74.91 & 21.82 & 2.6753 &  58.88 \\ 
        &  9:13 & 2330 &  80.74 & 33.63 & 1.8057 &  46.00 \\
        & 12:24 & 3554 & 113.44 & 74.31 & 1.0387 &   9.10 \\
        & 14:03 & 3569 & 130.77 & 79.15 & 1.0182 &  25.15 \\
\hline
18 Jun 2021 & 12:00 & 3641 & 105.04 & 69.39 & 1.0684 &  10.77 \\
        & 12:58 & 3711 & 136.40 & 80.39 & 1.0142 &  12.15 \\
        & 13:56 & 3747 & 224.12 & 80.32 & 1.0145 &  23.57 \\
        & 15:11 & 3060 & 259.23 & 65.71 & 1.0972 &  40.46 \\
        & 15:32 & 2312 & 263.29 & 61.23 & 1.1409 &  45.28 \\
\hline
19 Jun 2021 &  7:23 &  697 &  68.98 & 10.67 & 5.2698 & 71.34 \\
        &  7:59 & 1297 &  72.93 & 18.01 & 3.2067 & 63.11 \\
        &  9:06 & 2264 &  79.93 & 32.06 & 1.8839 & 47.71 \\
        & 10:27 & 3074 &  88.85 & 49.44 & 1.3163 & 29.24 \\
        & 11:17 & 3371 &  95.88 & 60.22 & 1.1522 & 18.39 \\
        & 11:46 & 3518 & 101.46 & 66.40 & 1.0912 & 12.90 \\
        & 13:08 & 3873 & 148.20 & 81.70 & 1.0106 & 13.79 \\
        & 14:12 & 3426 & 237.29 & 77.68 & 1.0236 & 27.05 \\
        & 15:15 & 2720 & 260.06 & 64.91 & 1.1042 & 41.33 \\
        & 15:46 & 1993 & 265.57 & 58.27 & 1.1757 & 48.45 \\
        & 16:45 &  900 & 273.28 & 45.55 & 1.4009 & 62.01 \\
        & 17:14 &  674 & 276.50 & 39.31 & 1.5785 & 68.65 \\
        & 17:35 &  501 & 278.72 & 34.82 & 1.7514 & 73.43 \\
\hline
23 Jun 2021 & 7:03 &  324  & 66.59  &  6.49 & 8.2685 & 76.09 \\
        & 7:29 &  744  & 69.57  & 11.69 & 4.8333 & 70.18 \\
        & 7:59 & 1266  & 72.85  & 17.82 & 3.2389 & 63.32 \\
        & 8:32 & 1773  & 76.33  & 24.69 & 2.3838 & 55.73 \\
        & 9:06 & 2237  & 79.85  & 31.87 & 1.8941 & 47.92 \\
\hline
26 Jun 2021 & 7:03 &  299  & 66.59 &  6.32 & 8.4534 & 76.25 \\
        & 7:25 &  665  & 69.13 & 10.72 & 5.2457 & 71.24 \\
        & 8:23 & 1684  & 75.41 & 22.64 & 2.5843 & 57.95 \\
        & 8:46 & 2050  & 77.80 & 27.47 & 2.1606 & 52.66 \\
        & 9:19 & 2554  & 81.24 & 34.47 & 1.7668 & 45.07 \\
        & 9:44 & 2944  & 83.90 & 39.82 & 1.5615 & 39.33 \\
\hline
21 Aug 2021 &  8:01 & 1002 &  83.66 & 12.81 & 4.4345 & 64.43 \\
          &  8:02 & 1016 &  83.78 & 13.02 & 4.3652 & 64.19 \\
          &  8:18 & 1342 &  85.74 & 16.46 & 3.4931 & 60.32 \\
          &  8:40 & 1739 &  88.46 & 21.20 & 2.7483 & 55.00 \\
          &  8:59 & 2050 &  90.87 & 25.30 & 2.3301 & 50.40 \\
          &  9:00 & 2054 &  91.00 & 25.52 & 2.3119 & 50.16 \\
          &  9:20 & 2325 &  93.64 & 29.83 & 2.0046 & 45.31 \\
          &  9:40 & 2850 &  96.43 & 34.13 & 1.7822 & 40.45 \\
          & 11:47 & 3916 & 123.06 & 60.00 & 1.1547 &  9.72 \\
\hline
\enddata
\tablenotetext{a}{Column by column we give the date,
the Central Daylight Time in hours and minutes, the measured
power (in W) from the solar panels, the Sun's azimuth and
elevation angle in degrees, the Sun's air mass, and the
angle in degrees between the vector perpendicular to the solar panels
and the direction toward the Sun.}
\tablenotetext{b}{Something is odd about the data of 13 June.
We derive the extinction term using the first four points only.}
\end{deluxetable}

\begin{deluxetable}{ccrrrrc}
\tabletypesize{\scriptsize}
\tablecolumns{7}
\tablewidth{0pc}
\tablecaption{Raw Data - Part II\tablenotemark{a}\label{tab:moredata}}
\tablehead{ \colhead{Date} &
\colhead{CST/CDT} &
\colhead{P$_{meas}$} &
\colhead{AZ$_{\odot}$} &
\colhead{EL$_{\odot}$} &
\colhead{X$_{\odot}$} &
\colhead{$\theta$} }
\startdata
 30 Oct 2021 &  8:48 & 1852 & 115.10 & 12.92 & 4.3990 & 56.82 \\
             &  9:18 & 2356 & 119.77 & 18.66 & 3.1004 & 50.51 \\
             & 10:01 & 2883 & 127.40 & 26.40 & 2.2409 & 42.10 \\
             & 10:29 & 3135 & 133.17 & 31.01 & 1.9410 & 37.26 \\
             & 11:00 & 3301 & 140.28 & 35.60 & 1.7177 & 32.79 \\
             & 11:29 & 3411 & 147.86 & 39.28 & 1.5795 & 29.81 \\
             & 12:06 & 3476 & 158.82 & 42.87 & 1.4698 & 28.32 \\
             & 12:31 & 3469 & 166.94 & 44.46 & 1.4276 & 28.96 \\
             & 12:58 & 3383 & 176.13 & 45.32 & 1.4063 & 31.05 \\
             & 13:28 & 3298 & 186.50 & 45.17 & 1.4099 & 34.78 \\
             & 14:00 & 2889 & 197.23 & 43.75 & 1.4461 & 39.91 \\
             & 14:10 & 2793 & 200.44 & 43.05 & 1.4648 & 41.70 \\
             & 14:19 & 2728 & 203.25 & 42.33 & 1.4850 & 43.37 \\
             & 14:20 & 2719 & 203.56 & 42.24 & 1.4874 & 43.56 \\
             & 14:30 & 2531 & 206.58 & 41.33 & 1.5142 & 45.47 \\
             & 14:31 & 2507 & 206.88 & 41.23 & 1.5172 & 45.67 \\
             & 14:40 & 2457 & 209.50 & 40.32 & 1.5456 & 47.43 \\
             & 14:50 & 2346 & 212.32 & 39.21 & 1.5820 & 49.44 \\
             & 14:56 & 2295 & 213.95 & 38.50 & 1.6064 & 50.67 \\
             & 15:00 & 2247 & 215.03 & 38.01 & 1.6239 & 51.50 \\
             & 15:10 & 2067 & 217.63 & 36.73 & 1.6721 & 53.59 \\
\hline
 22 Jan 2022 &  8:04 & 1303 & 118.70 &  8.02 & 6.8597 & 61.20 \\
             &  8:24 & 1747 & 121.63 & 11.75 & 4.8115 & 57.17 \\
             &  8:44 & 2141 & 124.75 & 15.36 & 3.7304 & 53.30 \\
             &  9:04 & 2477 & 128.09 & 18.84 & 3.0735 & 49.60 \\
             &  9:25 & 2768 & 131.87 & 22.31 & 2.6201 & 45.98 \\
             &  9:45 & 2989 & 135.74 & 25.43 & 2.3198 & 42.82 \\
             & 10:04 & 3172 & 139.70 & 28.19 & 2.1104 & 40.16 \\
             & 10:28 & 3318 & 145.11 & 31.35 & 1.9222 & 37.37 \\
             & 10:53 & 3432 & 151.25 & 34.19 & 1.7793 & 35.29 \\
             & 11:52 & 3499 & 167.65 & 38.69 & 1.5997 & 34.50 \\
             & 12:30 & 3401 & 179.15 & 39.64 & 1.5676 & 37.13 \\
             & 13:02 & 3262 & 188.90 & 39.15 & 1.5839 & 40.91 \\
             & 13:37 & 2968 & 199.16 & 37.31 & 1.6496 & 46.26 \\
\hline
 28 May 2022 &  7:00 &  396 &  68.46 &  5.96 & 8.8830 & 75.93 \\
             &  7:20 &  733 &  70.80 & 10.01 & 5.5952 & 71.32 \\
             &  7:40 & 1089 &  73.07 & 14.11 & 4.0448 & 66.70 \\
             &  8:00 & 1451 &  75.28 & 18.26 & 3.1649 & 62.07 \\
             &  8:20 & 1784 &  77.45 & 22.46 & 2.6040 & 57.41 \\
             &  8:40 & 2063 &  79.61 & 26.69 & 2.2185 & 52.75 \\
             &  9:00 & 2334 &  81.78 & 30.95 & 1.9446 & 48.09 \\
             &  9:20 & 2581 &  83.98 & 35.23 & 1.7335 & 43.43 \\
\enddata
\tablenotetext{a}{Column headings are like those in Table \ref{tab:data}
except the local time is CDT for 30 October 2021 and 28 May 2022, but 
CST for 22 January 2022.}
\end{deluxetable}

\begin{deluxetable}{cccc}
\tablecolumns{4}
\tablewidth{0pc}
\tablecaption{Derived Calibration Values\tablenotemark{a}\label{tab:extvals}}
\tablehead{ \colhead{Date} &
\colhead{Extinction (mag/airmass)} &
\colhead{Slope (W)} &
\colhead{N} }
\startdata
13 Jun 2021  & 0.124 $\pm$ 0.058 & 3905 $\pm$ 357 & 4  \\
14 Jun 2021  & 0.183 $\pm$ 0.049 & 3599 $\pm$ 112 & 9  \\
16 Jun 2021  & 0.130 $\pm$ 0.021 & 3744 $\pm$ 76  & 13 \\
19 Jun 2021  & 0.138 $\pm$ 0.068 & 3759 $\pm$ 176 & 9  \\
23 Jun 2021  & 0.152 $\pm$ 0.014 & 3836 $\pm$ 62  & 5  \\
26 Jun 2021  & 0.173 $\pm$ 0.017 & 4139 $\pm$ 55  & 6  \\
21 Aug 2021  & 0.181 $\pm$ 0.044 & 4091 $\pm$ 152 & 9  \\
\hline
28 Oct 2021 & 0.030 $\pm$ 0.017 & 4138 $\pm$ 207  & 6  \\
29 Oct 2021 & 0.072 $\pm$ 0.008 & 4293 $\pm$  73  & 8  \\
30 Oct 2021 & 0.051 $\pm$ 0.008 & 4077 $\pm$ 101  & 10 \\
 5 Nov 2021 & 0.081 $\pm$ 0.008 & 4248 $\pm$  69  & 14 \\
14 Nov 2021 & 0.100 $\pm$ 0.013 & 4114 $\pm$  80  & 11 \\
23 Nov 2021 & 0.085 $\pm$ 0.009 & 4329 $\pm$  34  & 8  \\
21 Dec 2021 & 0.097 $\pm$ 0.009 & 4390 $\pm$  69  & 9  \\
\hline
 3 Jan 2022 & 0.067 $\pm$ 0.007 & 4531 $\pm$ 41  & 8  \\
16 Jan 2022 & 0.092 $\pm$ 0.008 & 4539 $\pm$ 46  & 10 \\
22 Jan 2022 & 0.093 $\pm$ 0.013 & 4526 $\pm$ 78  & 13 \\
29 Jan 2022 & 0.074 $\pm$ 0.007 & 4374 $\pm$ 51  & 14 \\
 5 Feb 2022 & 0.065 $\pm$ 0.008 & 4495 $\pm$ 76  & 10 \\
13 Feb 2022 & 0.054 $\pm$ 0.011 & 4299 $\pm$ 69  & 9  \\
19 Feb 2022 & 0.086 $\pm$ 0.011 & 4687 $\pm$ 87  & 3  \\
12 Mar 2022 & 0.091 $\pm$ 0.004 & 4666 $\pm$ 40  & 4  \\
26 Mar 2022 & 0.136 $\pm$ 0.019 & 4262 $\pm$ 75  & 7  \\
 9 Apr 2022 & 0.127 $\pm$ 0.012 & 4196 $\pm$ 46  & 7  \\
26 May 2022 & 0.109 $\pm$ 0.005 & 3838 $\pm$ 18  & 8  \\ 
27 May 2022 & 0.154 $\pm$ 0.033 & 3850 $\pm$ 63  & 7  \\
28 May 2022 & 0.120 $\pm$ 0.018 & 3892 $\pm$ 53  & 8  \\
\enddata
\tablenotetext{a}{``Slope'' is the scaling factor of cos($\theta$),
giving $P_ {extcorr}$, the power corrected to Sun at the zenith.
Angle $\theta$ is the angular difference
of the direction to the Sun and the vector perpendicular to the panels.
N is the number of points used to determine the extinction term 
and the slope.}
\end{deluxetable}

\clearpage

\figcaption[]
{View of the house from the southeast.  There are six solar panels on the upper roof
and seven on the lower roof.
\label{fig:house_panels}
}

\figcaption[]
{Measured power (in Watts) from the solar panels as a function of time of 
day.  Color coding: 13 June (magenta); 14 June (blue); 16 
June (cyan); 19 June (green); 23 June (orange); 26 June (red); 21 August (brown); 
9, 10, 15, 17, and 18 June (grey). The vertical dashed line indicates the time 
of day in June 2021 when the house begins casting a shadow on some of the 
panels in the lower set of seven.
\label{fig:raw_data}
}

\figcaption[]
{Measured power (in Watts) from the solar panels (Z-axis) as a function 
of azimuth of the Sun in degrees (X-axis) and the elevation angle of the
Sun in degrees (Y-axis).  We have excluded data obtained when there
was frost on the solar panels, and have also excluded data obtained
when shadows were being cast on some of the panels.
\label{fig:3d_raw_data}
}

\figcaption[]
{The plane parallel atmosphere model. Ground level is represented by the 
solid horizontal line.  The extent of the atmosphere is represented by the 
dashed line. A star at zenith angle $z$ is observed through a path length 
of atmosphere that is approximately equal to sec($z$) times the path 
length in the direction of the zenith.  This ratio of path lengths is 
called the ``air mass'' (X).
\label{fig:slab}
}

\figcaption[]
{The difference of the color-corrected instrumental $V$-band magnitudes of 
photometric standards {\em minus} the standardized $V$-band magnitudes
from \citet{lan92,lan07} versus the air mass value (secant of the zenith angle).
The data were taken with the 0.9-m telescope at Cerro Tololo Inter-American
Observatory by the author on 26 November 2005 (UT). The yellow triangle represents
a very blue star measured at high air mass.  As it is a 5-$\sigma$ outlier,
it has been eliminated from the fit.
\label{fig:nov26}
}

\figcaption[]
{{\em Top:} The measured power output of the solar panel system (in 
Watts) on 16 June 2021 as a function of the cosine of the angular 
distance between the vector perpendicular to the solar panels and the 
direction toward the Sun. The regression line clearly does not pass 
through the origin. {\em Bottom:} Same as top diagram, except the power 
has been corrected to what would have been measured if the Sun were at 
the zenith and the panels were tilted to the same values of angle 
$\theta$.
\label{fig:jun16}
}


\figcaption[]
{Derived values of the extinction term $k_{\lambda}$ vs. the day of the year.
\label{fig:extvals}
}

\figcaption[]
{Power (corrected to Sun at zenith)
vs. cos($\theta$) for 22 January and 28 May 2022.  
\label{fig:slope}
}

\figcaption[]
{Values of the power (corrected to Sun at zenith) vs. the extinction term
$k_{\lambda}$.  We have color coded the data according to three sets
of months.
\label{fig:corr}
}

\figcaption[]
{The median atmospheric extinction at Cerro Tololo and Las Campanas 
as a function of wavelength, based on 29 nights of observations by us
from 21 April 2001 to 21 December 2012. The filters used were Johnson
$U$, $B$, and $V$, Cousins $R$ and $I$, and Sloan $u$, $r$, and $i$.
The dashed line corresponds to the mean wavelength of functionality 
of solar panels with silicon-based photovoltaic cells.
\label{fig:extinction}
}

\figcaption[]
{Power (corrected to Sun at zenith)
vs. cos($\theta$) for 30 October 2021.  The raw data are given in
Table \ref{tab:moredata}.  The solid line is fitted to the blue
points.  As on many occasions, the solar panels underperform the regression
line past a certain time of day.  This is most likely due to the decrease in
efficiency of the panels when operated at higher temperature.  The time stamps
are Central Daylight Time.
\label{fig:oct30}
}

\clearpage

\begin{figure}
\plotfiddle{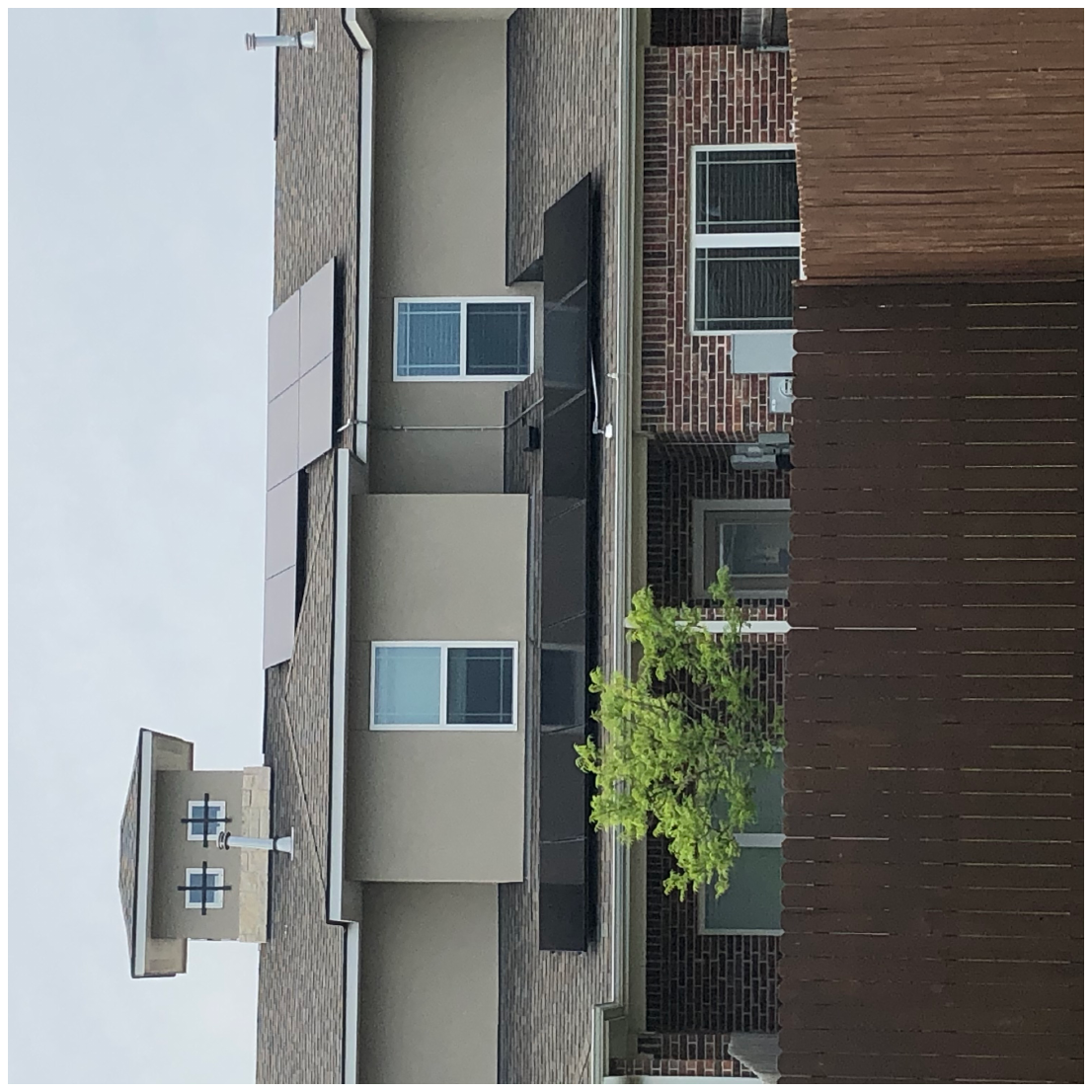}{3.5 in}{-90.0}{425}{425}{-50}{500}
\end{figure}

\begin{figure}
\plotone{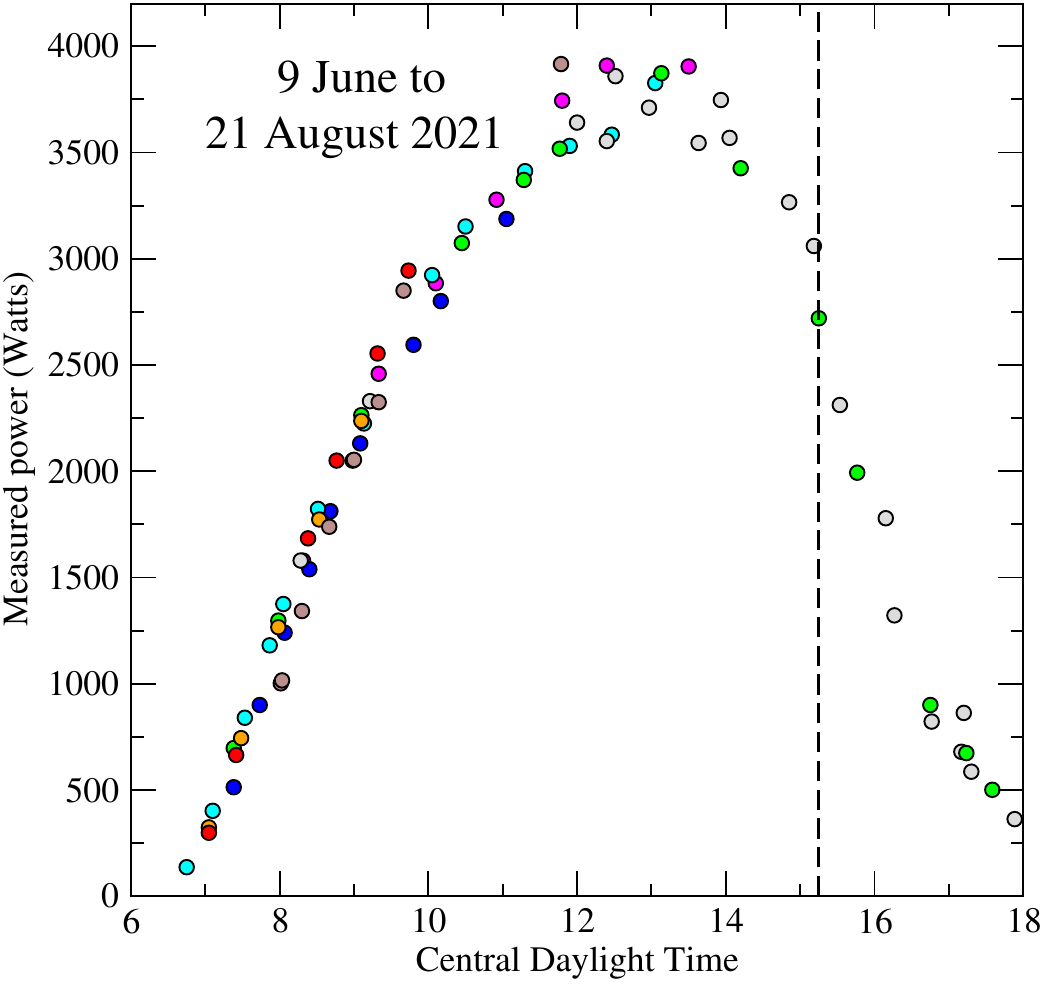} {Krisciunas Fig. \ref{fig:raw_data}. 
}
\end{figure}

\begin{figure}
\plotone{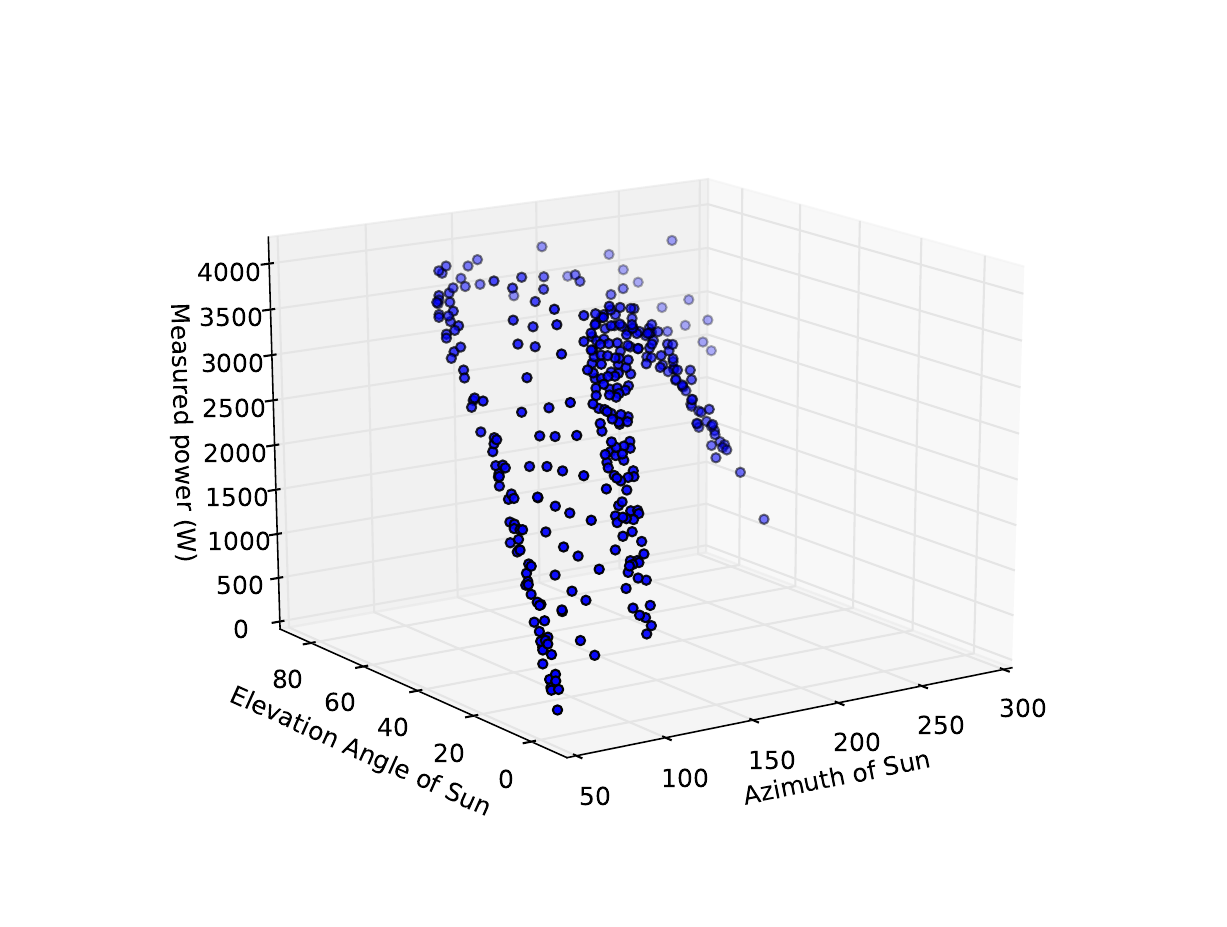} {Krisciunas Fig. \ref{fig:3d_raw_data}. 
}
\end{figure}

\begin{figure}
\plotone{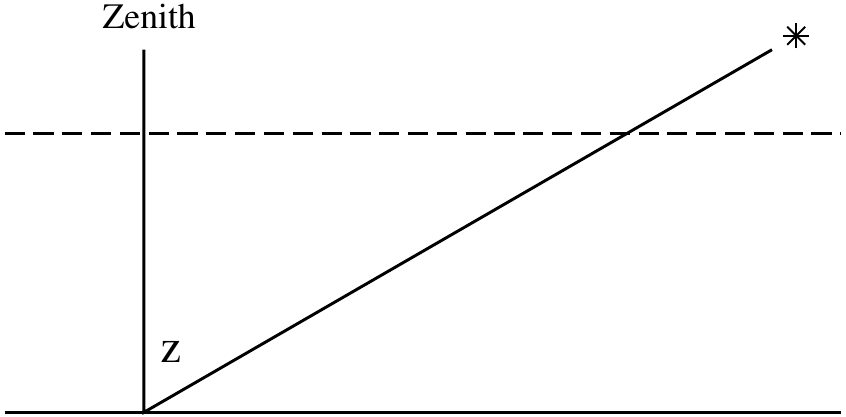} {Krisciunas Fig. \ref{fig:slab}. 
}
\end{figure}

\begin{figure}
\plotone{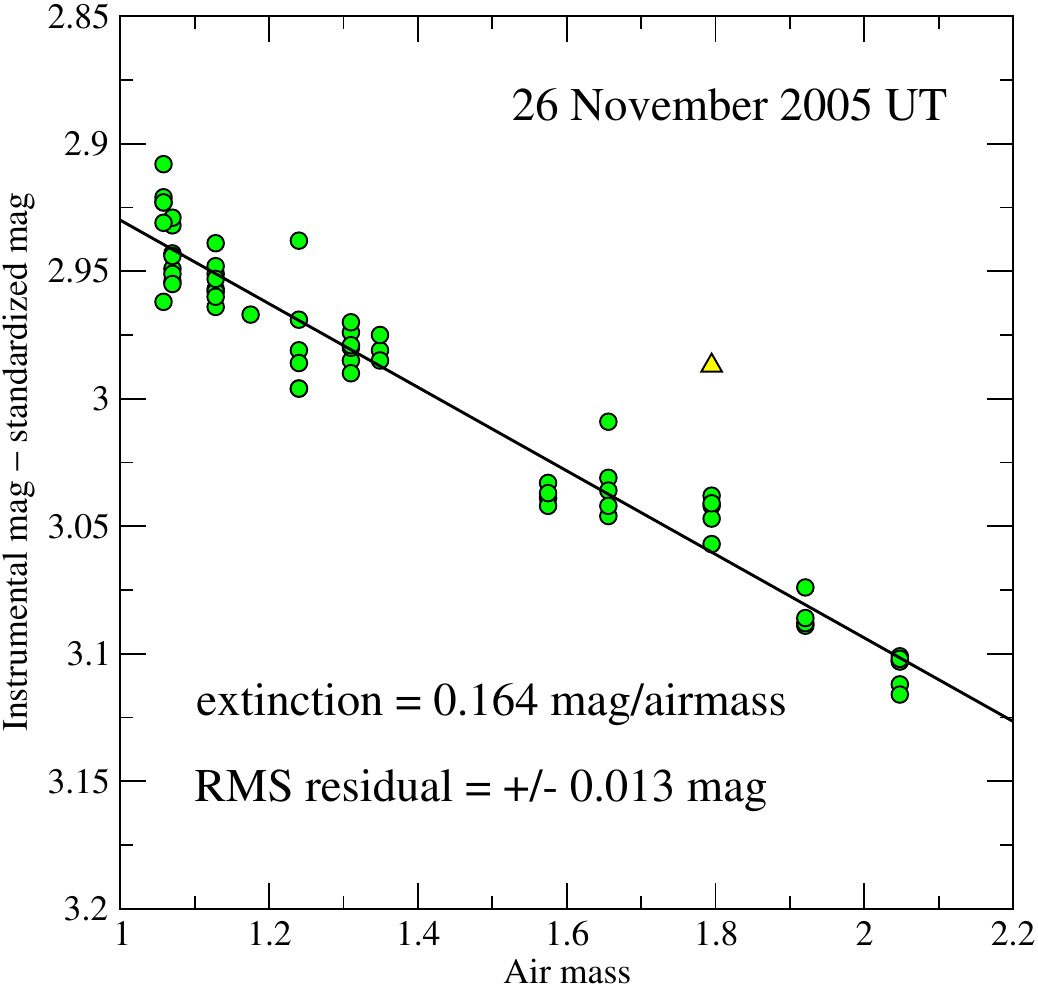} {Krisciunas Fig. \ref{fig:nov26}. 
}
\end{figure}

\begin{figure}
\plotone{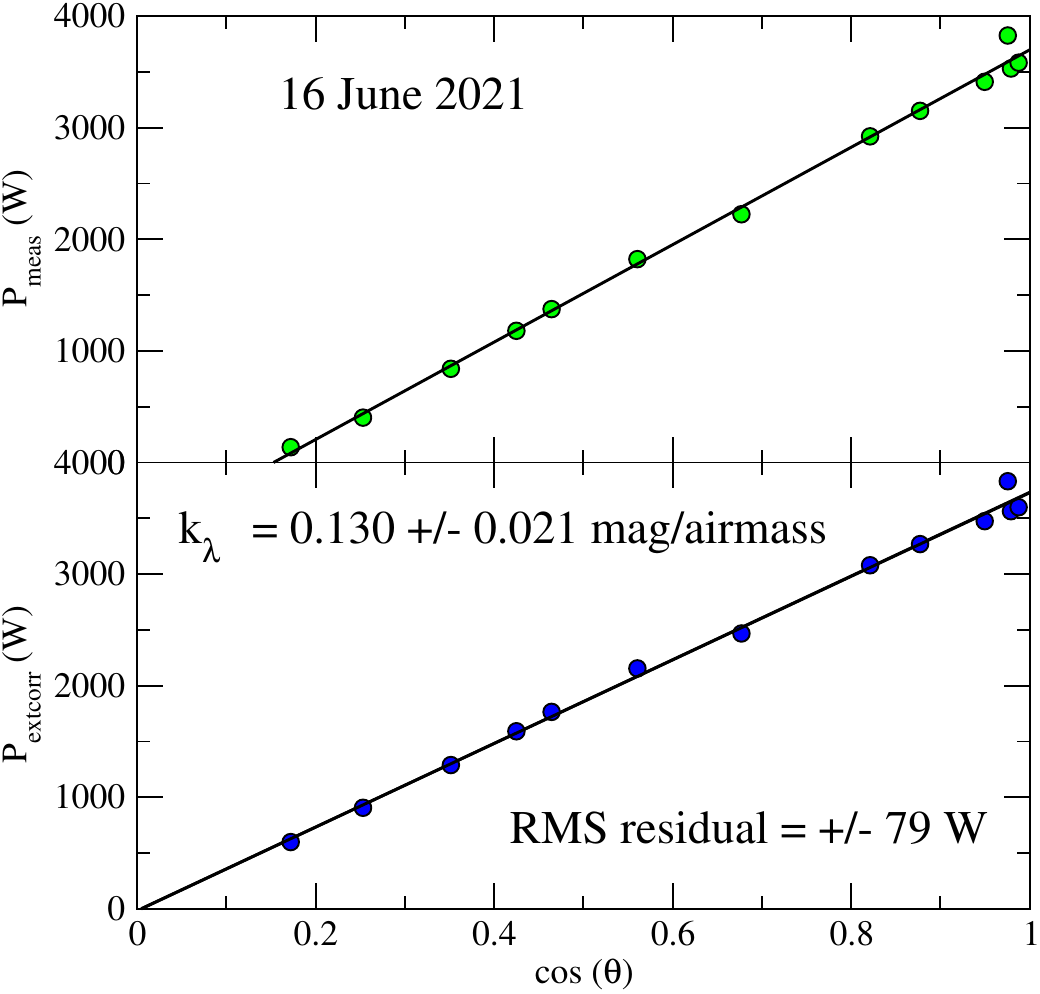} {Krisciunas Fig. \ref{fig:jun16}. 
}
\end{figure}

\begin{figure}
\plotone{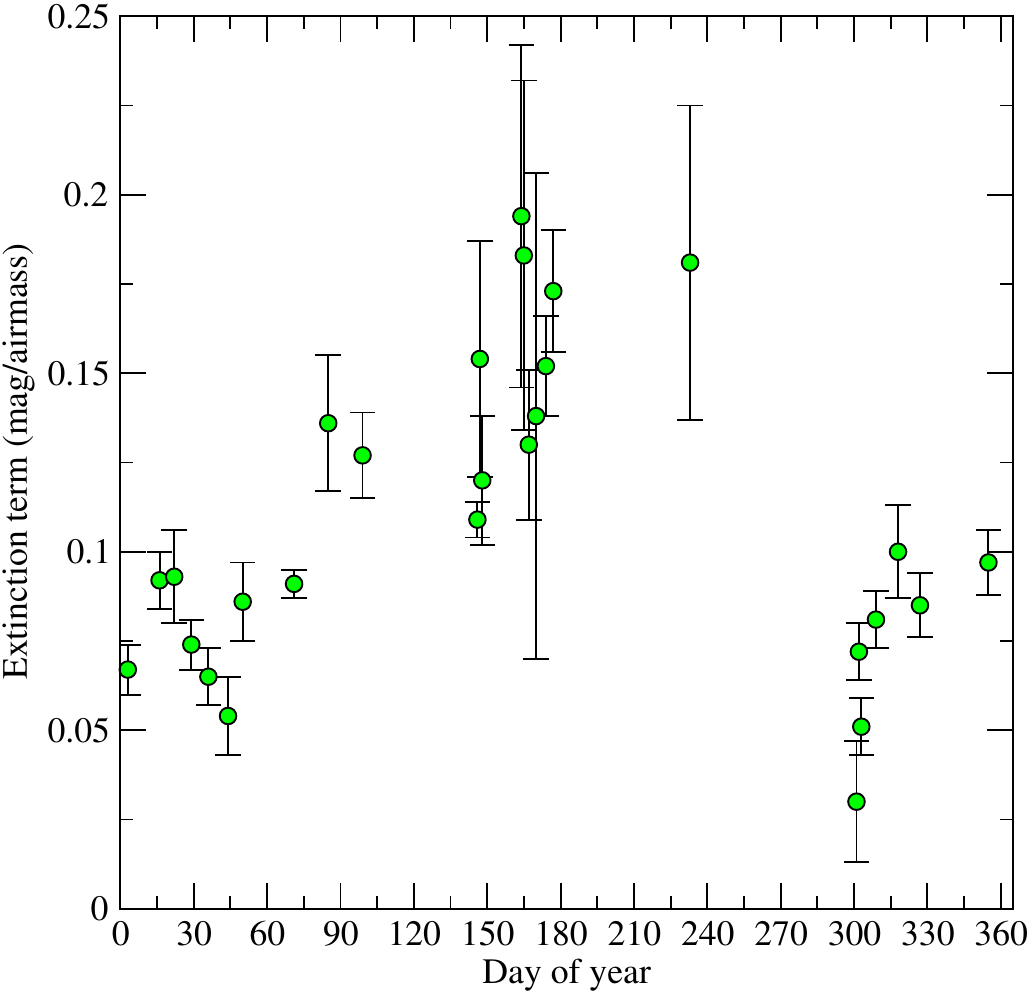} {Krisciunas Fig. \ref{fig:extvals}. 
}
\end{figure}

\begin{figure}
\plotone{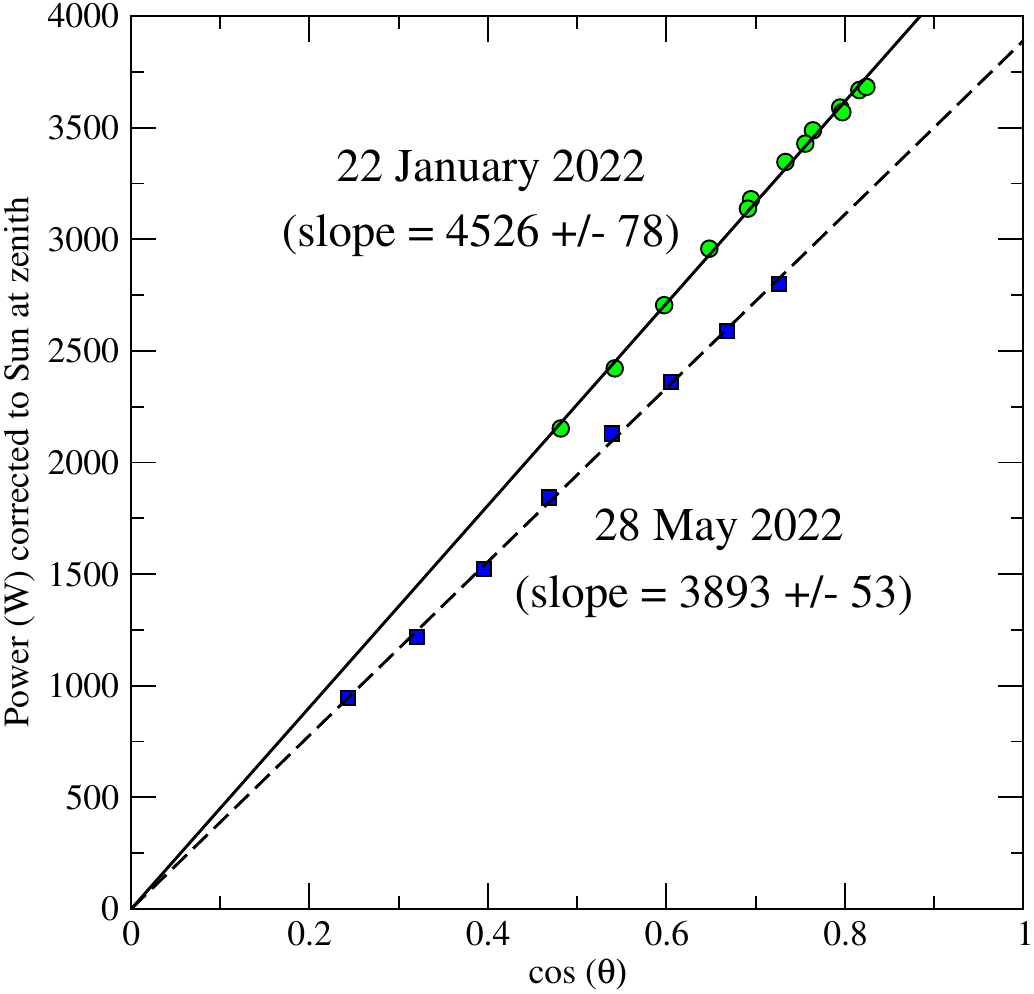} {Krisciunas Fig. \ref{fig:slope}. 
}
\end{figure}

\begin{figure}
\plotone{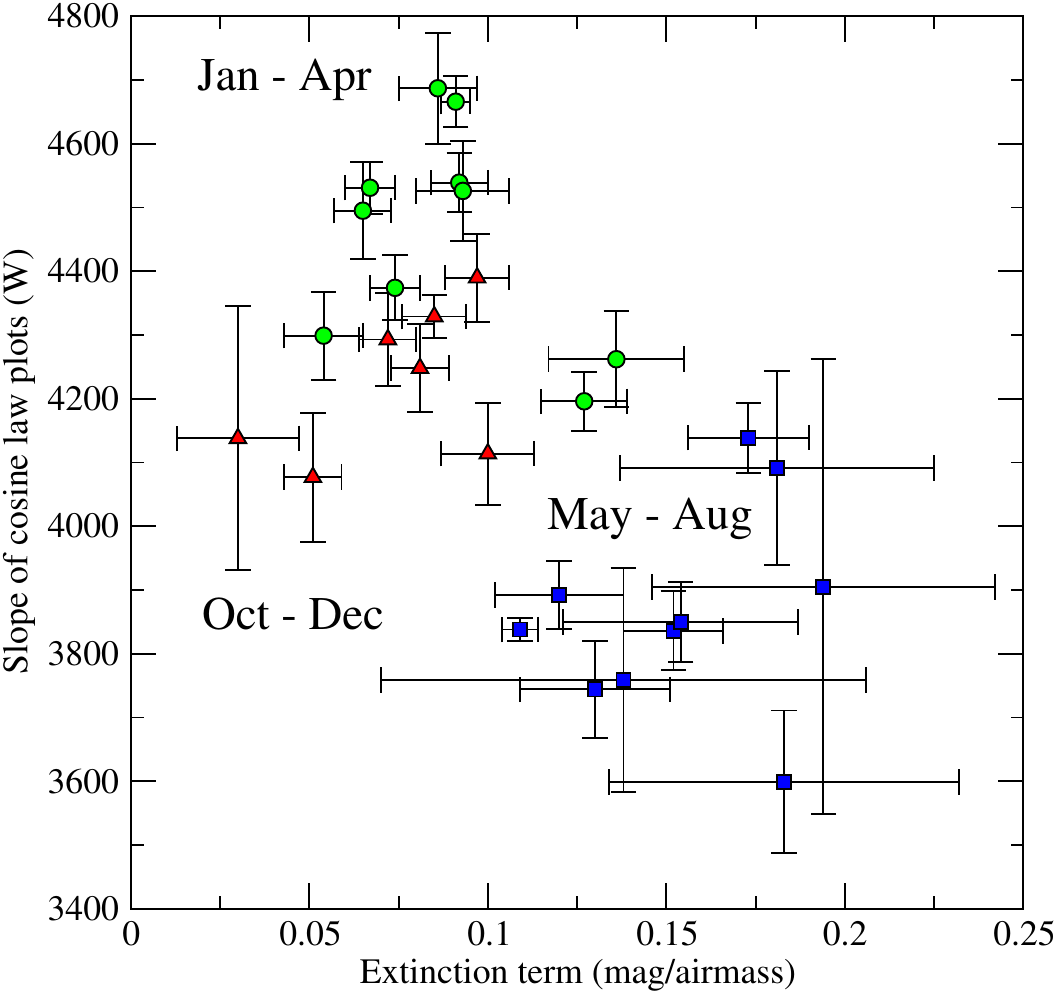} {Krisciunas Fig. \ref{fig:corr}. 
}
\end{figure}

\begin{figure}
\plotone{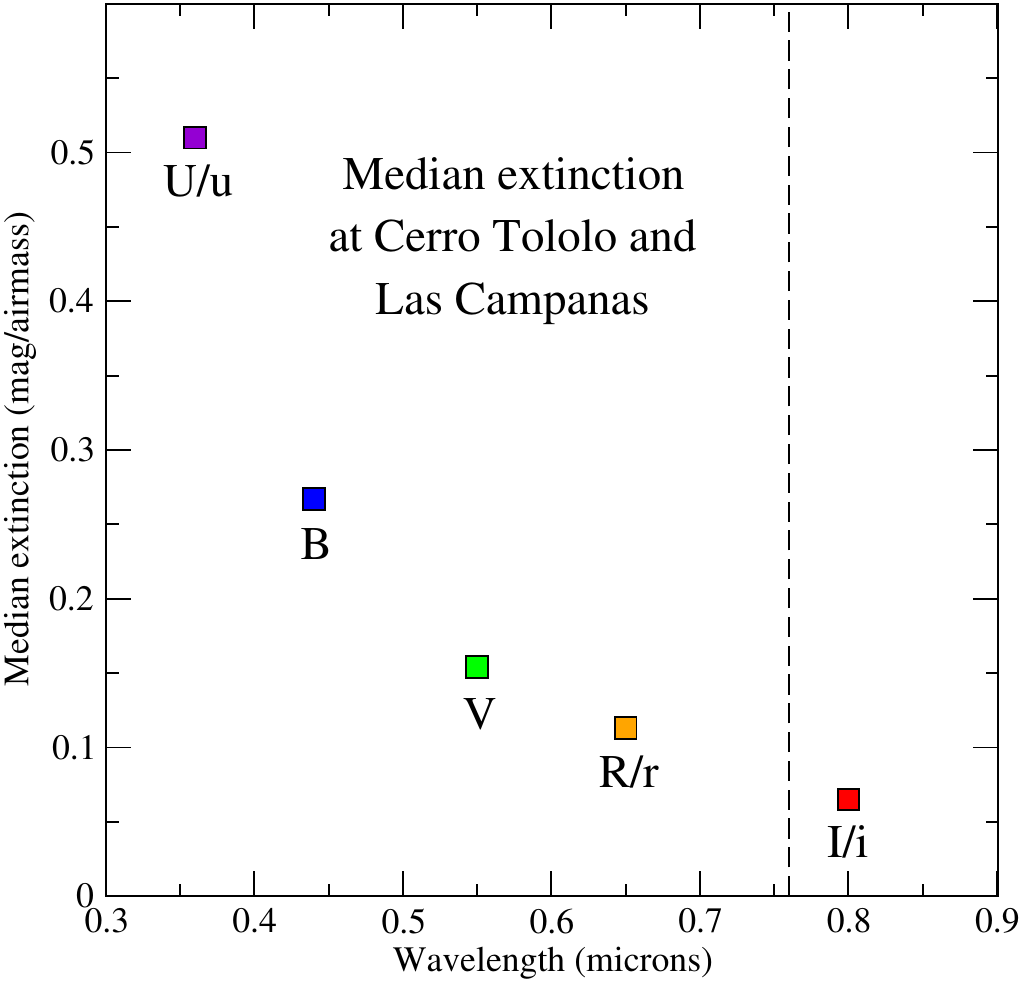} {Krisciunas Fig. \ref{fig:extinction}. 
}
\end{figure}

\begin{figure}
\plotone{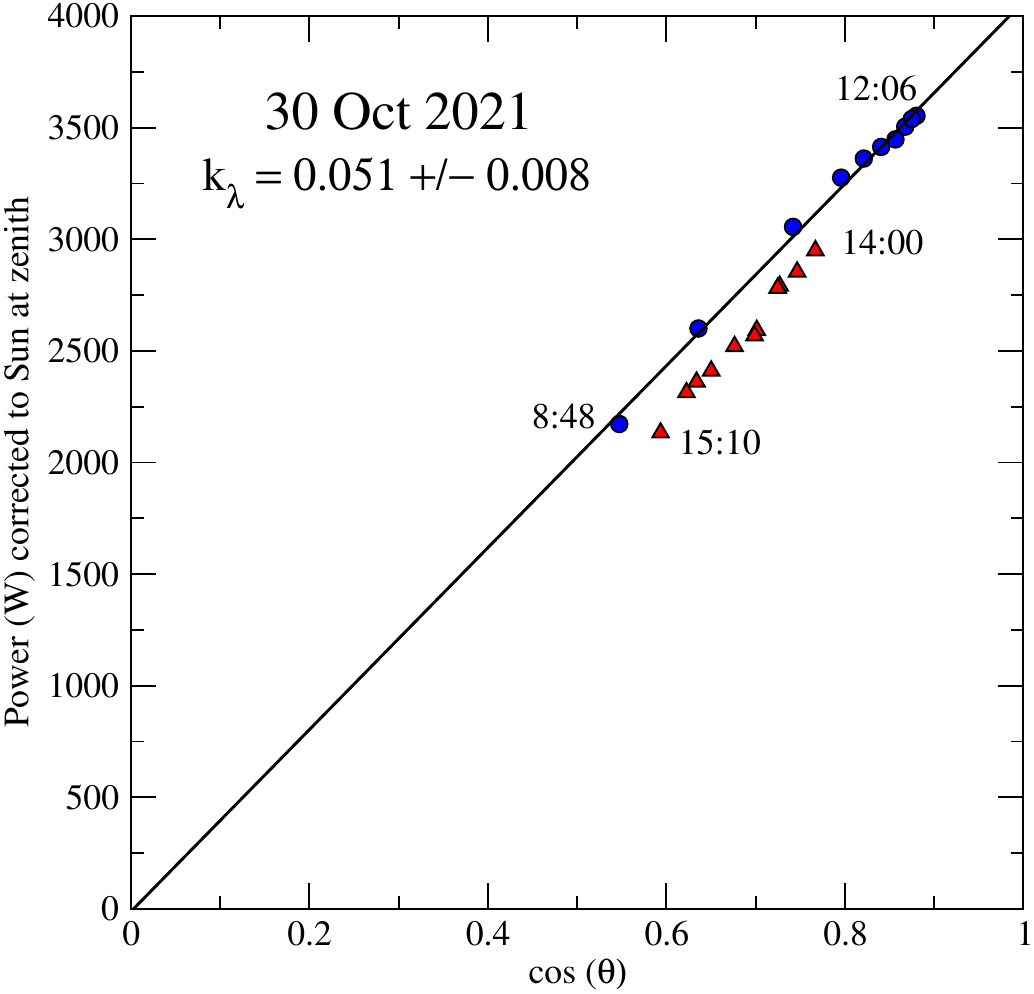} {Krisciunas Fig. \ref{fig:oct30}. 
}
\end{figure}

\end{document}